\documentclass{nature}
\usepackage{graphicx}
\usepackage{amsmath}
\usepackage{amsxtra}
\usepackage{amstext}
\usepackage{amssymb}
\usepackage{caption}
\newcommand{\ket}[1]{\ensuremath{|{#1}\rangle}}

\title{Creation and control of multi-phonon Fock states in a bulk acoustic wave resonator}

\author{Yiwen Chu$^{1*}$, Prashanta Kharel$^{1*}$, Taekwan Yoon$^1$, Luigi Frunzio$^1$, Peter T. Rakich$^1$, \& Robert J. Schoelkopf$^1$}

\begin{document}

\maketitle

\begin{affiliations}
 \item Department  of  Applied  Physics,  Yale  University,  New  Haven,  Connecticut  06511,  USA  and
Yale  Quantum  Institute,  Yale  University,  New  Haven,  Connecticut  06520,  USA
\end{affiliations}

\begin{abstract}

Quantum states of mechanical motion can be important resources for quantum information, metrology, and studies of fundamental physics. Recent demonstrations of superconducting qubits coupled to acoustic resonators have opened up the possibility of performing quantum operations on macroscopic motional modes\cite{OConnell2010, MooresPRL2017, ChuScience2017}, which can act as long-lived quantum memories or transducers. In addition, they can potentially be used to test for novel decoherence mechanisms in macroscopic objects and other modifications to standard quantum theory\cite{ArndtNatPhys2014, MarshallPRL2003}. Many of these applications call for the ability to create and characterize complex quantum states, putting demanding requirements on the speed of quantum operations and the coherence of the mechanical mode. In this work, we demonstrate the controlled generation of multi-phonon Fock states in a macroscopic bulk-acoustic wave resonator. We also perform Wigner tomography and state reconstruction to highlight the quantum nature of the prepared states\cite{LeibfriedPRL1996}. These demonstrations are made possible by the long coherence times of our acoustic resonator and our ability to selectively couple to individual phonon modes. Our work shows that circuit quantum acousto-dynamics (circuit QAD)\cite{Manenti2017} enables sophisticated quantum control of macroscopic mechanical objects and opens the door to using acoustic modes as novel quantum resources.

\end{abstract}

Light and sound are two familiar examples of wave phenomena in the classical world. By now, the field of quantum optics has extensively demonstrated the particle nature of light in quantum mechanics through the study of single photons and other non-Gaussian electromagnetic states. The concept of particles of sound, or phonons, is used widely in solid state physics. However, the ability to create states of individual phonons has only been demonstrated in a few instances\cite{OConnell2010, ChuScience2017, Riedinger2016}, while complete quantum tomography of such states has only been achieved in a single trapped ion\cite{LeibfriedPRL1996}. 
%the quantum state of a individual phonons has thus far only been created and measured in the motion of a single trapped ion\cite{LeibfriedPRL1996}. 
This disparity between electromagnetic and acoustic degrees of freedom is largely because sound propagates inside the complex and potentially lossy environment of a massive material rather than vacuum. As a result, an open question remains: Is it feasible to control and measure complex quantum states in the motion of a macroscopic solid state object, or what we usually think of as sound, analogously to what has been done with light? 

The relatively new field of quantum \textit{acoustics} is attempting to answer this question using a variety of optomechanical and electromechanical systems\cite{LeeScience2011, SafaviNaeini2013, Riedinger2018, OckeloenKorppi2018, OConnell2010, Gustafsson2014, Manenti2017, MooresPRL2017}, and one particularly promising approach within quantum acoustics is circuit QAD\cite{OConnell2010, Gustafsson2014, Manenti2017, MooresPRL2017}. In analogy to circuit quantum electro-dynamics (circuit QED), circuit QAD uses superconducting quantum circuits that operate at microwave frequencies to manipulate and measure mechanical resonators. Circuit QAD takes advantage of the strong interactions between mechanics and electromagnetism enabled by, for example, piezoelectricity. It also incorporates the non-linearity provided by the Josephson junction, which is a crucial ingredient for creating non-Gaussian states of motion. In turn, the ability to create these states make mechanical resonators useful as resources in quantum circuits, offering capabilities beyond those of electromagnetic resonators. For example, mechanical transduction is a promising method for transferring quantum information between microwave circuits and other systems such as optical light or spin qubits\cite{AndrewsNatPhys2013, Schuetz2015}. Due to the difference between the speeds of sound and light, an acoustic resonator is much more compact and well isolated than an electromagnetic one at the same frequency and provides many more independent modes that are individually addressable by a superconducting qubit. Such an architecture is desirable for simulating many-body quantum systems\cite{NaikNatComm2017, MooresPRL2017} and provides a highly hardware efficient way of storing, protecting, and manipulating quantum information using bosonic encodings\cite{LeghtasPRL2013, ChouArxiv2018}. These examples show that, by repurposing the toolbox of circuit QED through the similarities between light and sound, circuit QAD allows us to make use of the important differences between these quantum degrees of freedom. However, in order to access this toolbox, we first need to demonstrate that a circuit QAD system can be engineered to have the necessary mode structure, strong enough interactions, and sufficient quantum coherence to create and characterize quantum states of motion.

In this work, we experimentally prepare and perform full quantum tomography on Fock states of phonons and their superpositions inside a high-overtone bulk acoustic wave resonator (HBAR). 
%These demonstrations are enabled by making significant improvements to the performance of the system demonstrated in our previous work \cite{ChuScience2017}.
These demonstrations are enabled by a robust new flip-chip device geometry that couples a superconducting transmon qubit to the HBAR. This geometry allows us to separately optimize the design of the acoustic resonator and qubit to extend phonon coherence while enhancing the selectivity of the coupling to a single mode. The combination of these improvements leads to a device that is deeper in the strong coupling regime of circuit QAD, which is necessary for the generation and manipulation of more complex quantum states. We note that a similar demonstration using a superconducting qubit and surface acoustic waves was recently reported\cite{Satzinger2018}.

% without sacrificing the robustness of the fabrication process. 
% Specifically, we separate the design and characterization of phonon and qubit modes\cite{Satzinger2018}, increase the coherence of our acoustic resonator, and improve our ability to selectively couple to one phonon mode. 
We now motivate and describe the design of our circuit QAD system in more detail. Figure 1a shows a schematic of our device, which we call the $\hbar$BAR from now on. The first important difference from our previous device is the flip-chip geometry, where the qubit and acoustic resonator are now on separate sapphire chips\cite{Satzinger2018}. This simplifies the fabrication procedure and increases the yield of successful devices (see Supplementary Information), while allowing for qubits and acoustic resonators to be individually tested before assembly. Second, the $\hbar$BAR now incorporates a plano-convex acoustic resonator that is fabricated using a simple, robust recipe and supports stable, transversely confined acoustic modes (see Supplementary information). Since the measured acoustic lifetime in the previous unstable resonator geometry\cite{ChuScience2017} was consistent with being limited by diffraction loss, this modification to our device could significantly improve the phonon coherence.
%  that ideally do not suffer from the diffraction loss that limited the phonon lifetime of our previous device. 
%The convex surface is fabricated using a simple and robust recipe\cite{Kharel2018} that can produce a range of controllable radii of curvature (see Supplementary information). 
%Another significant limitation of our previous device was the simultaneous coupling to multiple transverse acoustic modes. 
Another important requirement is the ability to selectively couple the qubit to a single acoustic mode. This is partly achieved by the plano-convex resonator design, which allows us to control the frequency spacing between transverse modes. To further increase mode selectivity, the third improvement is the addition of an optimized transduction electrode to the qubit. The electrode was designed to match the strain profile of the fundamental Gaussian transverse mode of the acoustic resonator (see Supplementary information). We point out that even though the acoustic resonator is not in physical contact with the electrode, the electric field of the qubit extends across the gap between the two chips and through the AlN film, thus allowing for piezeoelectric transduction. 

We now experimentally show that the new design does indeed lead to improvements in the electro-mechanical coupling, acoustic mode spectrum, and coherence of our device. As in our previous work, the $\hbar$BAR is measured using a standard circuit QED setup that allows for flux tuning of the qubit frequency. Figure 1b shows qubit spectroscopy near the $l = l_1$ and $m, n = 0$ mode of the $\hbar$BAR, which reveals a single distinct anticrossing feature (Figure 1b). Here $l$ is the longitudinal mode number, and $m, n$ are the mode numbers of the Hermite-Gaussian-like transverse modes. $l_1\sim466$ corresponds to the highest frequency longitudinal mode fully within the tunable range of the qubit, as indicated in Figure 2, where we investigate the mode structure of the $\hbar$BAR over several longitudinal free-spectral ranges. Figure 2a shows time dynamics of the qubit-phonon interaction, which reveals vacuum Rabi oscillations every $\nu_{\textrm{FSR}} = 13.5$ MHz as we tune the qubit frequency, each corresponding to an anticrossing feature similar to the one shown in Figure 1b. The Fourier transform of the data in Figure 2a is shown in Figure 2b and gives a qubit-phonon coupling rate of $g_0 = 2\pi \times (350 \pm 3)$ kHz. In addition to the dominant set of  oscillations corresponding to the $m, n = 0$ Gaussian modes, there are clear signatures of other acoustic modes visible in Figures 2a and b, which simulations indicate correspond to higher order transverse modes (see Supplementary Information). However, the closest observable higher order mode is $\sim$1 MHz away from the $m, n = 0$ mode and about ten times less strongly coupled to the qubit, while all others are at least five times less strongly coupled. From now on, we use only the longitudinal mode number to represent the $m, n = 0$ modes. These results indicate that the $\hbar$BAR is a good approximation of a system in which the qubit can be tuned to interact with a single acoustic mode at a time. 

We demonstrate improvements in the coherence of our system by performing quantum operations on the phonon mode using the qubit.
% To measure the phonon coherence, we use techniques described in our previous work\cite{ChuScience2017} that combine standard procedures for measuring qubit coherence and qubit-phonon swap operations that involve bringing the two systems on resonance for a time $T_s = \pi/2 g_0$. 
Using techniques described in our previous work\cite{ChuScience2017}, we find that the phonon mode has a $T_1$ of $(64\pm2)$ $\mu$s, a Ramsey $T_2$ of $(38\pm2)$ $\mu$s, and an echo $T_2$ of $(45\pm2)$ $\mu$s. On other devices, we measured that the phonon $T_1$ can be as long as $(113\pm4)$ $\mu$s. These coherence times are now comparable to that of state-of-the art superconducting qubits and suggest that the plano-convex resonator design does indeed support much longer lived phonons. The qubit in this device has a $T_1$ of $(7\pm1)$ $\mu$s, which is similar to our previous device. As will be discussed later, we believe these device parameters can be further improved through modifications of the materials, fabrication procedure, and device geometry.

The improvements presented above allow us to perform quantum operations on the phonon mode with a new level of sophistication, which we now illustrate by creating and measuring multi-phonon Fock states. We use a procedure for Fock state preparation that has previously only been demonstrated in electromagnetic systems\cite{Hofheinz2008} (Figure 3a). The experiment begins with the qubit set to a frequency $\nu_0$ that is $\delta = -5$ MHz detuned from the target $l_1$ phonon mode at frequency $\nu_1$. The qubit ideally starts out in the ground state $\ket{g}$, but in reality has a thermal population of $4-8\%$ in the excited state $\ket{e}$. The phonon modes, on the other hand, were shown to be colder\cite{ChuScience2017}. Therefore we first perform a swap operation between the qubit and the $l_2$ mode with frequency $\nu_2$. This procedure effectively uses an additional acoustic mode to cool the qubit to an excited state population of $\sim 2\%$. The qubit is then excited with a $\pi$ pulse and brought into resonance with the $l_1$ mode to swap the energy into the acoustic resonator. This is repeated $N$ times to climb up the Fock state ladder, ideally resulting in a state of $N$ phonons, which is then probed by bringing the qubit and phonon on resonance for a variable time $t$ and measuring the final qubit state. We note that this measurement procedure gives the total population in the qubit excited state subspace of the joint system and traces over the resonator state. The resulting time dynamics of $p_{e, N}(t)$ for up to $N=7$ are shown in Figure 3b. In Figure 3c, we plot the Fourier transform of the data in Figure 3b. As expected, we observe oscillations with a dominant frequency of $2g_N = 2\sqrt{N}g_0$, corresponding to the rate of energy exchange between the $\ket{g, N}$ and $\ket{e, N-1}$ states. 

In order to more quantitatively characterize the states we have created, we extract the population in each phonon Fock state $n$ after performing a $N$ phonon preparation. We do this by first simulating the expected time traces $p_{e, n}(t)$ if the phonon mode is prepared in an ideal Fock state ranging from $n=1$ to $n_{\textrm{max}}=14$. The independently measured value of $g_0$, along with the qubit and phonon decay and dephasing rates, are used in the simulations.
%, which result in $n_{\textrm{max}}$ time traces $p_{e, n}(t)$. 
Then, the experimental data for each $N$ (Figure 3b) are fitted to a weighted sum of the form
\begin{equation}
p_{e, N}(t) = \sum_{n=1}^{n_{\textrm{max}}}p_{n, N} p_{e, n}(t),
\end{equation}
where $p_{n, N}$ is then the population in $\ket{g, n}$ after performing a $N$ phonon preparation. The fit for each $N$ is subject to the constraints
$p_{n, N}\leq1\ \forall n$ and  $\sum_{n=1}^{n_{\textrm{max}}}p_{n, N} \leq 1$. Finally, the population in the zero phonon state is calculated as $p_{0, N} = 1-\sum_{n=1}^{n_{\textrm{max}}}p_{n, N}$.  Ideally, $p_{n, N} = \delta_{n, N}$. As shown in Figure 3d, we observe that the resulting distribution of populations for each experiment is indeed peaked at $n=N$. However, the population in the nominally prepared state decreases with increasing $N$. We find that $p_{1, 1} = 0.86$ , which is consistent with a simple estimate taking into account the energy decay from the one excitation manifold during a swap operation, which is dominated by the qubit decay rate, and the imperfect preparation of the qubit in $\ket{g}$. For larger $N$'s, the state preparation may be affected by additional effects such as off-resonant driving of the phonon mode during the qubit $\pi$ pulses, which could lead to excess population in the $n>N$ states. We also found that the largest source of potential error in extracting $p_{n, N}$ comes from uncertainty in the system parameters that are used in simulating $p_{e, n}(t)$. In particular, slight drifts of the qubit frequency can result in a mismatch between the value of $2g_0$ used in the simulations and the actual oscillation frequency of the vacuum Rabi data. An estimate of the effect of such miscalibrations are given by the errorbars in Figure 3d. 

We now build upon our ability to extract the phonon number distribution to perform full Wigner tomography and explore the quantum nature of the prepared mechanical state. As in previous experiments in circuit QED and trapped ions\cite{LeibfriedPRL1996, Hofheinz2009}, we make use of the definition\cite{RoyerPRA1977}
\begin{equation}
P(\alpha) = \textrm{Tr}[\hat{D}(-\alpha)\rho\hat{D}(\alpha)\hat{P}] = \frac{\pi}{2}W(\alpha).
\end{equation}
Here $P(\alpha)$ and $W(\alpha)$ are the values of the displaced parity and Wigner functions at a phase space amplitude $\alpha$, respectively, $\rho$ is the prepared state, and $\hat{P}$ is the parity operator. From now on we will plot the values of $P(\alpha)$ for clarity, but use the terms displaced parity and Wigner function interchangeably. The resonator displacement $\hat{D}(\alpha)$ is implemented by a microwave pulse at the phonon frequency while the qubit is detuned at $\nu_0$. Under these conditions, the phonon mode is still coupled to the microwave drive port, in part due to its hybridization with the qubit. To verify this and calibrate our displacement amplitudes, we first apply a Gaussian phonon drive pulse of varying amplitude $\alpha$ with a 1 $\mu$s RMS width and truncated to 4 $\mu$s total length. We then measure the subsequent Fock state populations $p_{n, \ket{0}}(\alpha)$ and check that they agree well with the expected Poisson distributions up to an overall scaling between the amplitudes of the applied drive and the actual displacement (see Supplementary Information). We can then calculate the displaced parity for the vacuum state $\ket{0}$ using $P_{\ket{0}}(\alpha) = \sum_n (-1)^n p_{n, \ket{0}}(\alpha)$. Similarly, we can measure the displaced parity $P_{\rho}(\alpha)$ for an arbitrary state $\rho$ by adding a phonon drive pulse between state preparation and measurement. 

In Figure 4, we present the results of Wigner state tomography on the nominally prepared states $\ket{1}$, $(\ket{0}+\ket{1})/\sqrt{2}$, and $\ket{2}$. The $(\ket{0}+\ket{1})/\sqrt{2}$ state was prepared by performing a $\pi/2$ pulse on the qubit and followed by a swap operation with the phonon mode. From the measured data shown in Figures 4a, b, and c, we can reconstruct the measured state using a maximum likelihood method\cite{ChouArxiv2018} (see Supplementary Information). The Wigner functions of the reconstructed states are presented in Figures 4d, e, and f. Figures 4g, h, and i show that the reconstructed parities agree well with the raw data. The negativity of the Wigner functions clearly demonstrate the quantum nature of the states.  From the reconstructed density matricies, we find that the fidelities of the prepared states to the target states are $F_{\ket{1}} = 0.87 \pm 0.01$, $F_{(\ket{0}+\ket{1})/\sqrt{2}} = 0.94 \pm 0.01$, and $F_{\ket{2}} = 0.78 \pm 0.02$. The infidelity for all three states are dominated by excess population in the lower number Fock states, which is an expected consequence of energy decay during state preparation and measurement (see Supplementary Information). 

These results show that the quantum state of motion in a macroscopic mechanical resonator can be prepared, controlled, and fully characterized in a circuit QAD device. The demonstration of even more complex quantum states should be possible with further improvements of the device performance. Currently, the dominant source of loss is the qubit, and we found that its $T_1$ is higher when the resonator chip is either not present or rotated by 180$^\circ$ relative to the qubit chip. This indicates that the qubit lifetime may be limited by loss due to the AlN, which could be mitigated by using a different piezoelectric material or optimizing the device geometry to minimize the electric field in the AlN that does not contribute to transduction. The current limitations on the phonon coherence also require further investigation. The energy loss is likely to be dominated by surface roughness or imperfections in the fabricated geometry, while additional dephasing could result from thermal excitations and frequency fluctuations of the detuned qubit\cite{GambettaPRA2006}. In addition, we can more carefully characterize the final flip-chip geometry, such as the spacing and alignment between the chips. The assembly process can then be modified accordingly, potentially leading to further improvements in the coupling and mode selectivity. 

The next generation of devices could give us access to even more sophisticated methods for quantum control of the acoustic resonator. Our current device is close to being able to reach the strong dispersive regime where circuit QED systems currently operate, which would allow for quantum non-demolition measurements of phonon numbers\cite{SchusterNature2007} and more sophisticated techniques for generating arbitrary quantum states of harmonic resonators\cite{HeeresNatComm2017, LeghtasPRL2013}. Furthermore, our technique for cooling the qubit already takes advantage of the multimode nature of the acoustic resonator. Future experiments would, for example, demonstrate qubit-mediated interactions between multiple modes and the creation of multipartite entangled states of mechanical motion\cite{Riedinger2018, OckeloenKorppi2018}. Recent efforts in improving the efficiency of electromechanical and optomechancial transduction with mechanical resonators could enable conversion of quantum information between the microwave and optical domains\cite{AndrewsNatPhys2013, Kharel2018}. Beyond the use of acoustic resonators as resources for quantum information, the creation of increasingly complex quantum states in highly coherent mechanical resonators can provide insight into the question of whether quantum superpositions of massive objects are suppressed due to mechanisms other than environmental decoherence\cite{ArndtNatPhys2014, Penrose1996}. In addition, the ability to perform quantum control on our large effective mass, high frequency, and low thermal occupation mechanical system may put new bounds on modifications to quantum mechanics at small length scales\cite{PikovskiNatPhys2012, Marin2013}. These examples suggest that the wide range of quantum acoustics demonstrations that may soon be possible with $\hbar$BAR will give rise to new quantum technologies while furthering our understanding of fundamental physics. 

\bibliography{QPv2Arxiv.bib}

\begin{addendum}
 \item We thank Michel Devoret, Steve Girvin, Yaxing Zhang, Kevin Chou, and Vijay Jain for helpful discussions.We thank Katrina Silwa for providing the Josephson parametric converter amplifier. This research was supported by the US Army Research Office (W911NF-14-1-0011), ONR YIP (N00014-17-1-2514), NSF MRSEC (DMR-1119826), and the Packard Fellowship for Science and Engineering. Facilities use was supported by the Yale SEAS cleanroom, the Yale West Campus Cleanroom, and the Yale Institute for Nanoscience and Quantum Engineering (YINQE).
 \item[Author contributions] Y.C. performed the experiment and analyzed the data under the supervision of P.T.R. and R.F.S. Y.C., P.K., and L.F. designed and fabricated the device. P.K. and T.Y. provided experimental suggestions and theory support. Y.C., P.K., P.T.R, and R.J.S. wrote the manuscript with contributions from all authors.
 \item[Author information]: Reprints and permissions information is available at www.nature.com/reprints. R.J.S., and L.F. are founders and equity shareholders of Quantum Circuits, Inc. Correspondence and requests for materials
should be addressed to Y. Chu (email: yiwen.chu@yale.edu) or R. J. Schoelkopf (robert.schoelkopf@yale.edu ).

% \item[Competing Interests] R.J.S., and L.F. are founders and equity shareholders of Quantum Circuits, Inc.
% \item[Correspondence] Correspondence and requests for materials
%should be addressed to Y. Chu (email: yiwen.chu@yale.edu) or R. J. Schoelkopf~(robert.schoelkopf@yale.edu ).
\end{addendum}

 \newpage

%\begin{figure}[H]
\begin{center}
%\begin{singlespace}
\includegraphics[width = 0.5\textwidth]{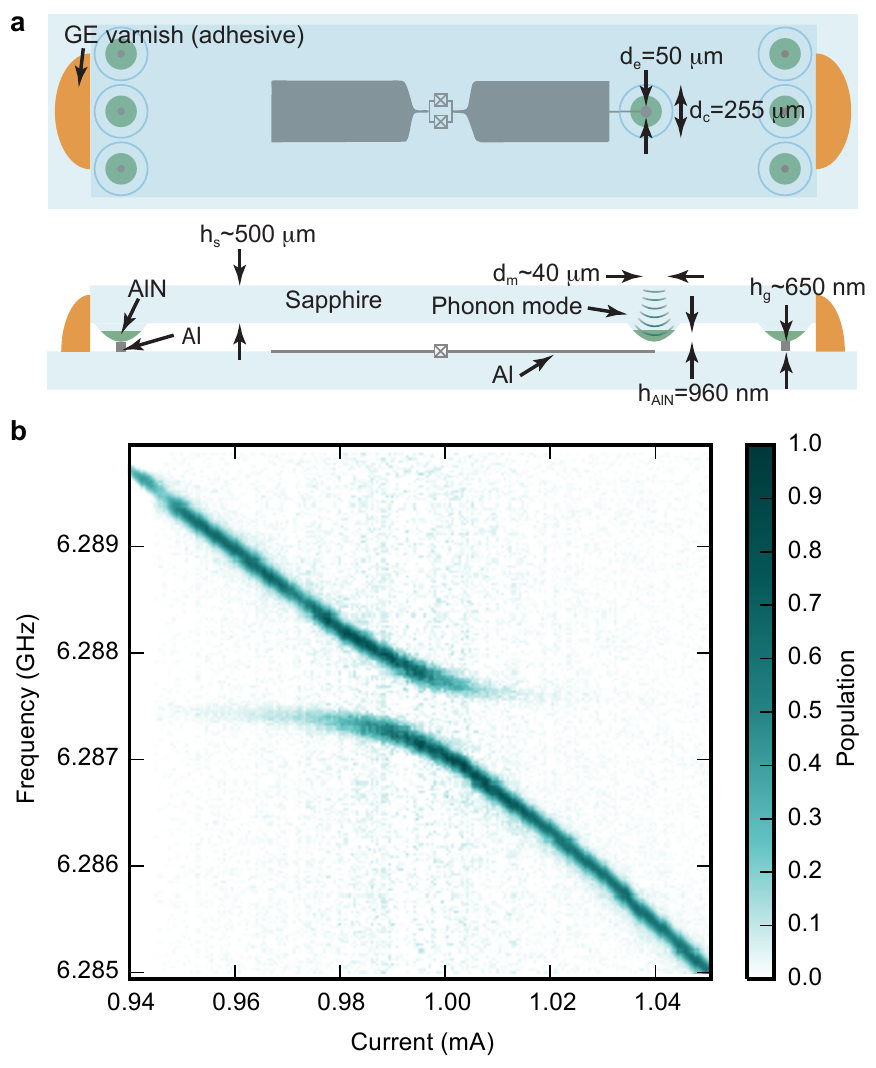}
\captionof{figure}{
\textbf{The $\hbar$BAR and strong qubit-phonon coupling. a}, Top and side view schematic of the $\hbar$BAR (not to scale). The chip containing the acoustic resonator also has six nominally identical resonators on the edges of the chip that have Al spacers deposited on them. The measured thickness of the Al spacer between the qubit and acoustic resonator ($h_g$) is shown, but the actual spacing may be larger due to imperfections in the flip-chip assembly. Other dimensions indicated are the diameters of the transducer electrode ($d_e$), curved resonator surface ($d_c$), and acoustic mode waist ($d_m$), along with the thicknesses of the AlN ($h_{\textrm{AlN}}$) and sapphire substrate ($h_s$). \textbf{b}, Spectroscopy of the transmon qubit near the $l_1$ acoustic mode while varying the current in an external coil used to flux tune the qubit frequency. 
}
%\label{Fig1}
%\end{singlespace}
\end{center}
%\end{figure}

 \newpage

%\begin{figure}[H]
\begin{center}
\includegraphics[width = \textwidth]{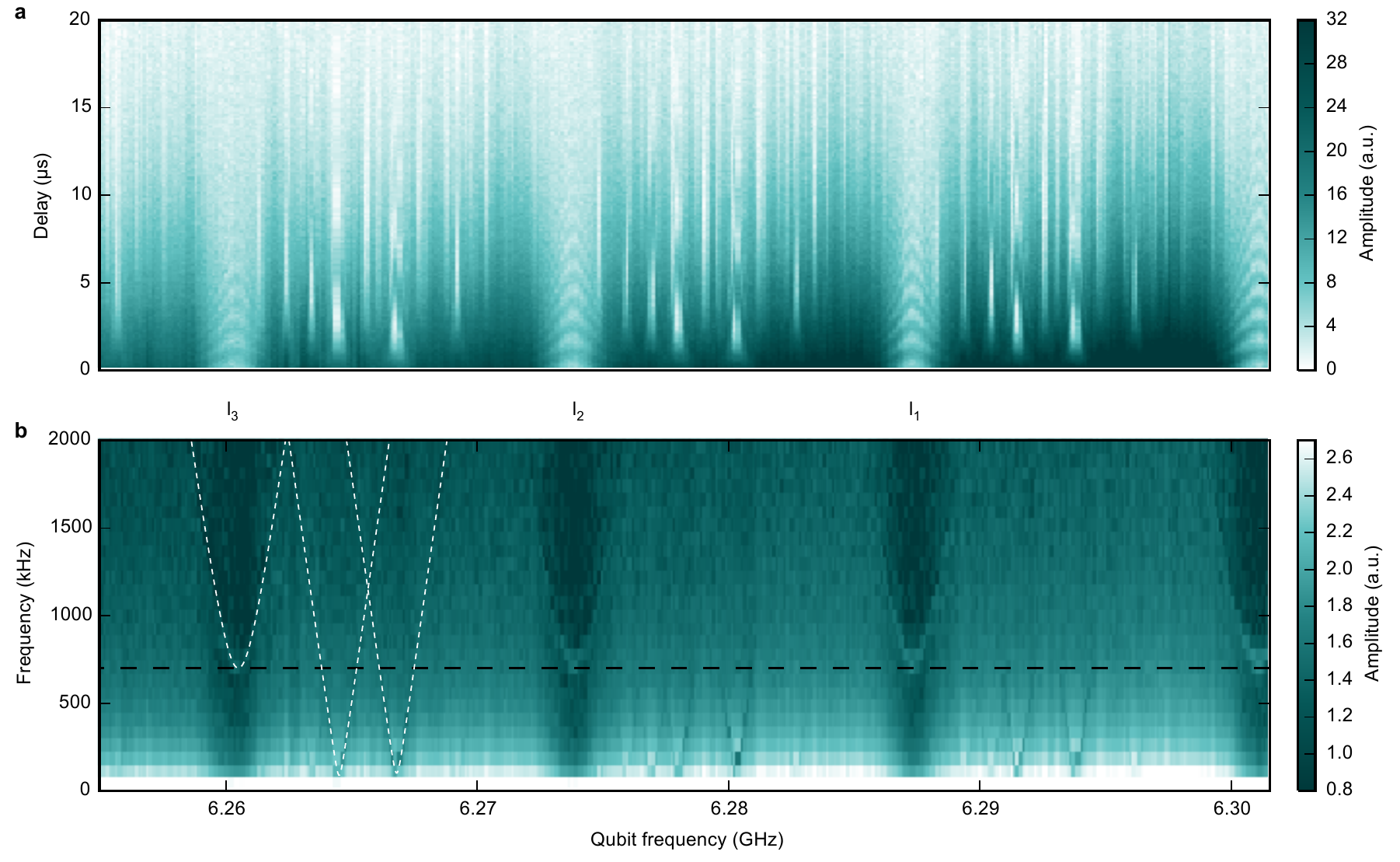}
\captionof{figure}{\textbf{Mode structure of the $\hbar$BAR. a}, Result of exciting the qubit and measuring its excited state population after a variable delay as the qubit is flux tuned. \textbf{b}, Logarithm of the Fourier transform of the data in a. The qubit frequencies shown on the horizontal axis are determined from spectroscopy data taken at each applied flux. The three highest frequency fundamental transverse modes ($m, n = 0$) that are fully accessible by the qubit are shown and labeled with their longitudinal mode numbers. The white dashed lines indicate the hyperbolic dependence of the effective vacuum Rabi frequencies for the three most dominant modes in one FSR. The black dashed line indicates the value of $2g_0$ for the fundamental mode, which is at least a factor of five larger than the coupling rates to the other modes. }
%\label{Figure2}
\end{center}
%\end{figure}
 \newpage
 
%\begin{figure}[H]
\begin{center}
\includegraphics[width = \textwidth]{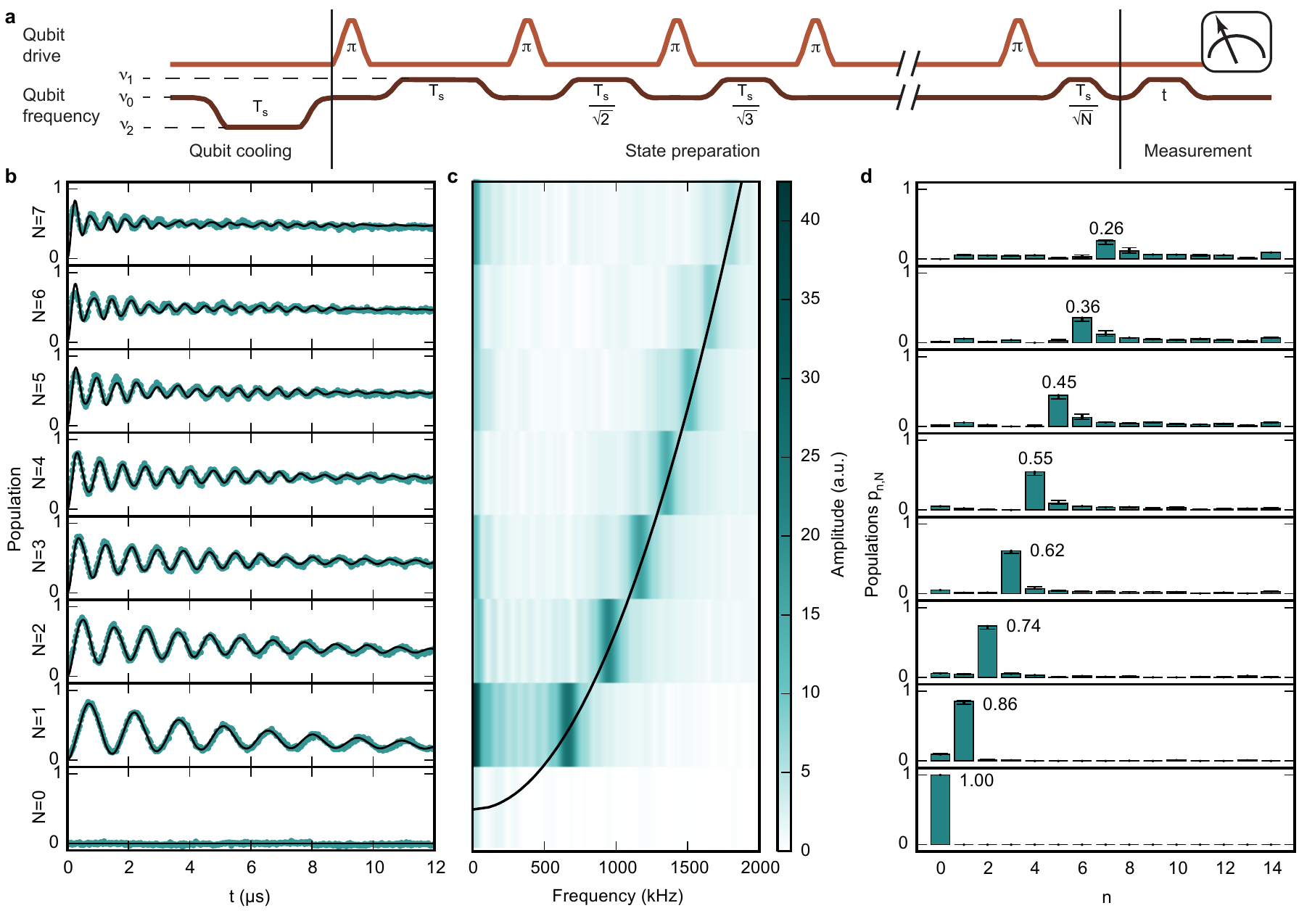}
\captionof{figure}{\textbf{Climbing the phonon Fock state ladder. a}, Pulse sequence for the generation and measurement of phonon Fock states. $T_s = \pi/2 g_0$ is the duration of a swap operation in the one excitation manifold. In the state preparation step, the duration of the $k$th swap is scaled to account for the coupling rate of $g_k = \sqrt{k}g_0$ between the $\ket{g, k}$ and $\ket{e, k-1}$ states. Pulses intended to excite the qubit from $\ket{g}$ to $\ket{e}$ are labeled with $\pi$. Qubit frequencies $\nu_0$, $\nu_1$, and $\nu_2$ are described in the text. \textbf{b}, Qubit excited state population after interacting with the phonon for a time $t$ following a $N$ phonon preparation procedure. Black lines are fits used to extract the Fock state populations shown in d. \textbf{c}, Fourier transform of the data in b, obtained by subtracting the mean of each dataset in b and padding with the resultant final value to effectively smooth the Fourier transform. Black line is $2g_N$. \textbf{d}, Populations in Fock state $n$ extracted from b. Numbers show the populations in $n=N$. Error bars indicate the result of changing the value of $g_0$ used in the simulations of $p_{e, n}(t)$ by $\pm5$ kHz.}
%\label{Figure2}
\end{center}
%\end{figure}
 \newpage
%\begin{figure}[H]
\begin{center}
\includegraphics[width = \textwidth]{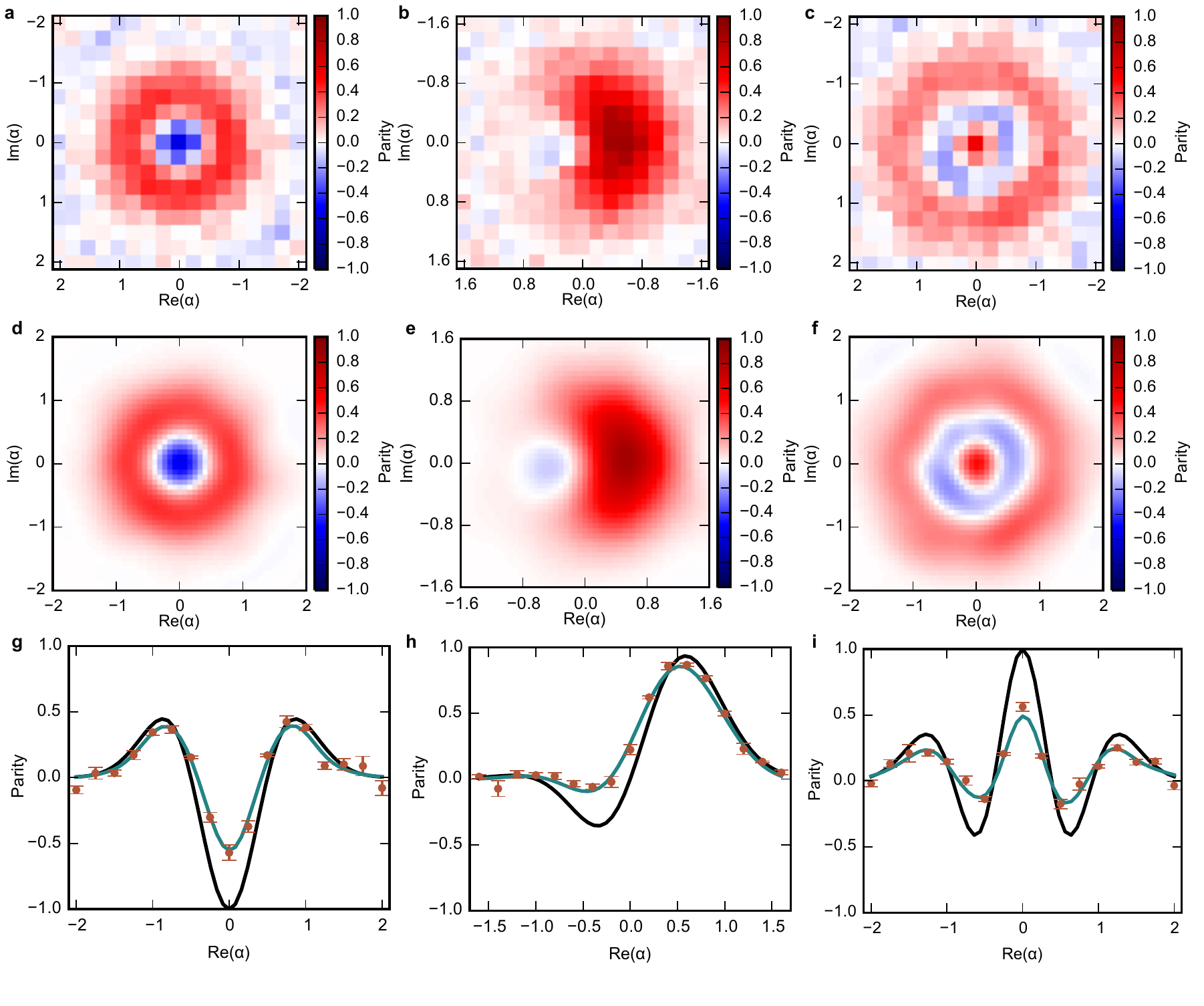}
\captionof{figure}{\textbf{Wigner tomography of non-classical states of motion. a, b, c}, Measured Wigner functions of the prepared states $\ket{1}$, $(\ket{0}+\ket{1})/\sqrt{2}$, and $\ket{2}$. Each grid point is a separate experiment with displacement by a phase space amplitude $\alpha$. \textbf{d, e, f}, Wigner functions of the density matrices reconstructed from a, b, and c. \textbf{g, h, i}, Cuts of the ideal Wigner function (black line), data (brown points), and reconstructed Wigner function (green line) along the Im$(\alpha)=0$ axis. Negative values of the Wigner function are indicators of a non-classical state of motion. Error bars on the data are extracted in the same way as in Figure 3d. 
}
%\label{Figure2}
\end{center}
%\end{figure}

 \newpage

\section*{Data availability}
The data that support the findings of this study are available from the corresponding 
authors upon reasonable request.

%%%%%%%%%%%
\clearpage

%%%%%%%%%%%%%%%%%%%%%%%%%%%
%%   SUPPLEMENTARY MATERIAL
%%%%%%%%%%%%%%%%%%%%%%%%%%%

\pagebreak
%\onecolumngrid
\begin{center}
\textbf{\large Supplementary information for:\\ Climbing the phonon Fock state ladder}
\end{center}

%%%%%%%%%% Prefix a "S" to all equations, figures, tables and reset the counter %%%%%%%%%%
\setcounter{equation}{0}
\setcounter{figure}{0}
\setcounter{table}{0}
\setcounter{page}{1}
\setcounter{section}{0}
\makeatletter
\renewcommand{\theequation}{S\arabic{equation}}
\renewcommand{\thefigure}{S\arabic{figure}}
\renewcommand{\thetable}{S\arabic{table}}

\section{Fabrication procedures}
\begin{figure}[ht] 
\begin{center}
\includegraphics[width = 0.9 \textwidth]{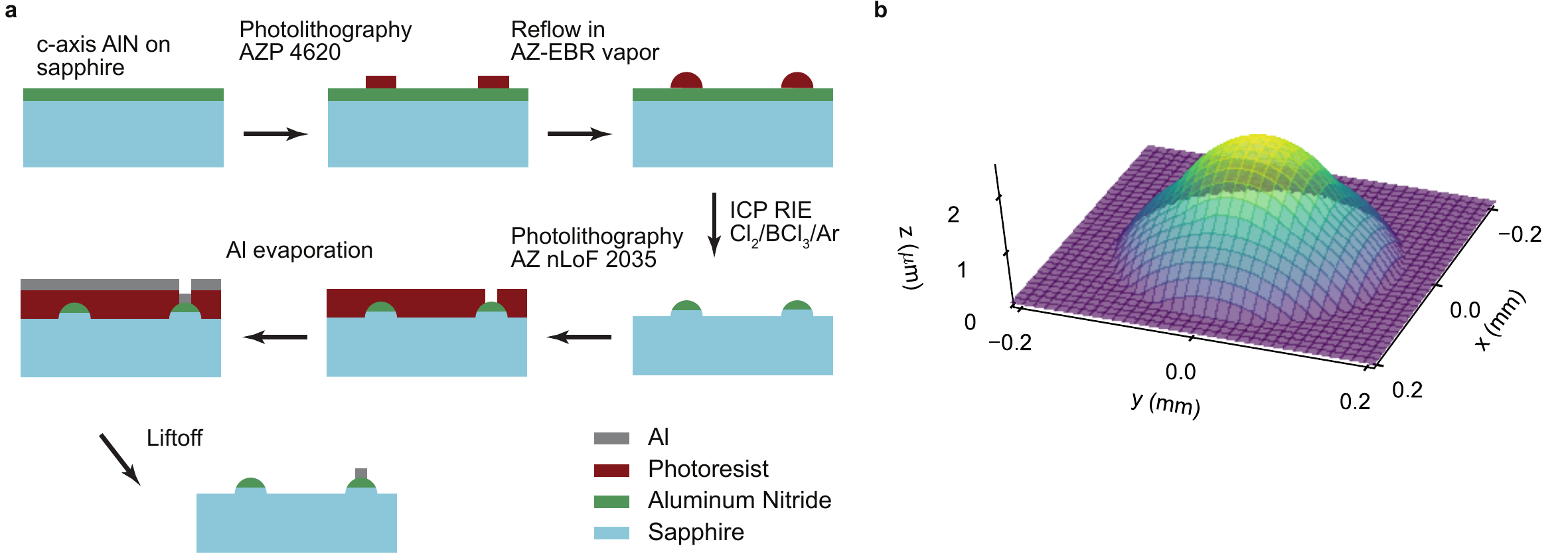}
\caption{\textbf{Fabrication of acoustic resonator chip. a.} Fabrication procedure. Drawings show a cross section through a part of the chip that includes the actual acoustic resonator device and one of the spacer resonators at the edges of the chip. \textbf{b.} Convex surface of acoustic resonator measured using optical profilometry.}
\label{fab}
\end{center}
\end{figure}

The transmon qubit used in our device is fabricated on sapphire using a standard e-beam lithography and Dolan bridge process\cite{Wang2015SI}. The fabrication procedures for the acoustic resonator chip is shown in Figure \ref{fab}a. We begin with a commercially purchased double side polished 2" sapphire wafer with a 1$\mu$m thick film of c-axis oriented AlN grown on one side (Kyma Technologies, part number H.AT.U.050.1000). The convex surfaces of the device resonator and the spacer resonators at the edges of the chip are then fabricated on the side with AlN using a procedure that has previously been demonstrated for making micro-lenses and high-Q high-overtone bulk acoustic wave resonators (HBAR) in other materials\cite{Kharel2018SI}. The procedure uses photoresist that has been reflown into hemispheres as a mask for reactive ion etching (RIE), which transfers the hemispherical geometry into the substrate due to the finite etch selectivity. 

We first pattern disks of photoresist (AZP 4620) on the wafer, which is then attached to a heated chuck at 55$^\circ$ C and inverted above a beaker of solvent (AZ-EBR) on a hotplate at 60$^\circ$ C. After 2-3 hours, the disks of resist will have reabsorbed the solvent and reflowed into hemispheres due to surface tension. The mode structure of the resonator depends on the radius of curvature of the final convex surface, which in turn depends on the radius of curvature and geometry of the reflowed resist hemisphere. The radii of each hemisphere is approximately that of the original disk, while the height and shape depend on the disk radius, the thickness of the resist, and the reflow time. Therefore, we use identical radii for the device and spacer resonators within each chip to ensure that they result in nominally identical heights of the final surface. 

After reflowing the resist, we bake the wafer starting at 90 $^\circ$C, gradually increasing to 145 $^\circ$C over the course of 15 minutes to remove the solvent and harden the resist. The wafer is then etched in an Oxford 100 RIE/ICP etcher (Cl$_2$/BCl$_3$/Ar at 4/26/15 sccm, 8 mTorr, 70 W RF power, 350 W ICP power). The resist mask has an etch rate of $\sim$200 nm/min, while AlN and sapphire have etch rates of $\sim$ 60 nm/min and $\sim$ 20 nm/min, respectively. We stop the etch when the maximum thickness of the AlN is $\sim\lambda/2$, where $\lambda$ is the acoustic wavelength at the qubit frequency, to maximize the qubit-phonon coupling strength. The different etch rates of the AlN and sapphire results in a two-tiered geometry of the final surface, as shown in Figure \ref{fab}b. However, the simulated mode radius is only $\sim$20 $\mu$m (see Section \ref{sims}), which means that the mode does not extend beyond the AlN part of the curved surface. 

After making the resonators, we use photolithography, e-beam evaporation, and liftoff to add Al spacers on top of the spacer resonators. The wafer is then diced into individual chips, which are then combined with the qubit chips using a home-built alignment and flip-chip bonding setup. We first pick up the resonator chip with a small drop of PDMS at the end of a tungsten tip, which is attached to a three-axis translation stage. We then use the stage to align the Al spacers on the resonator chip to corresponding features patterned on the qubit chip, which results in the device resonator being aligned with the transduction electrode on the qubit. It is also possible to directly align the device resonator using interference rings that are visible once the two chips are brought into contact. Once alignment is achieved, we add small drops of GE varnish on the edges to hold the chips together. 

\section{Acoustic mode simulations}\label{sims}

Detailed descriptions of the design and simulation of high-Q HBAR resonators are given in\cite{Renninger2018SI, Kharel2018SI, ChuScience2017SI}. The simulation propagates the acoustic wave over many roundtrips through the resonator and calculates the interferometric sum of the fields. At frequencies where a stable mode exists, the interference is constructive, and we obtain a well-defined mode profile with a large total intensity. We note that the simulated mode spectrum should not depend on the initial excitation, but certain modes can be missed if the initial excitation has particular symmetries. Therefore, we use an initial acoustic excitation that was found by first simulating the qubit electric field pattern with HFSS, then calculating the strain profile that is generated by this field through the piezoelectric transducer. 

\begin{figure}[ht] 
\begin{center}
\includegraphics[width = 0.8 \textwidth]{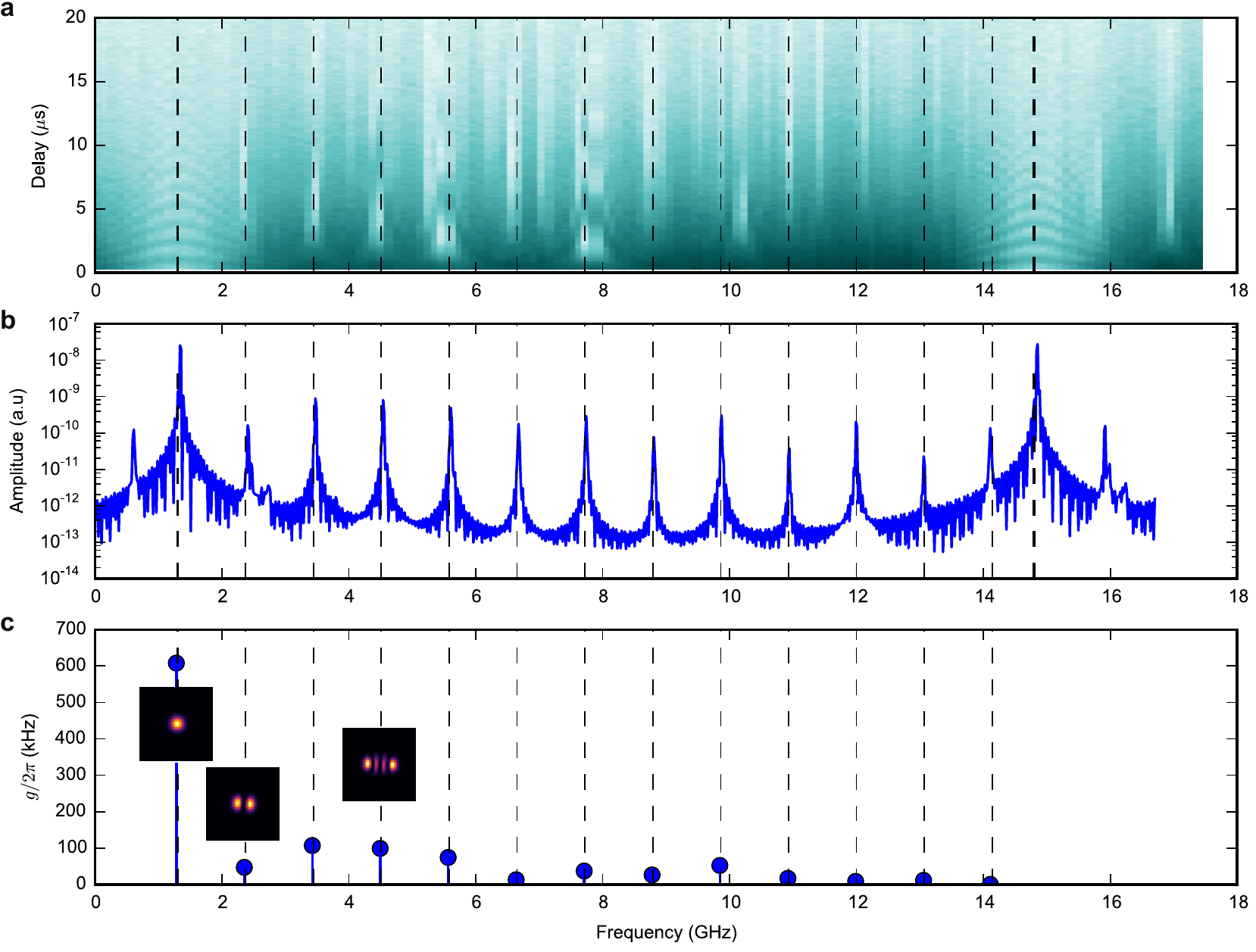}
\caption{\textbf{Acoustic resonator simulations. a.} Zoomed in view of one acoustic FSR from Figure 2a of the main text. \textbf{b.} Simulated acoustic mode spectrum. \textbf{c.} Simulated coupling strengths for a 1 $\mu$m spacing between the qubit and resonator chips. Insets show simulated strain profiles for three modes in a 300 $\mu$m $\times$ 300 $\mu$m area. Thick dashed lines indicate the locations of the fundamental Gaussian modes, while thin dashed lines indicate the locations of higher order Hermite-Gaussian modes. }
\label{modeSims}
\end{center}
\end{figure}

We now show that the simulated mode structure and coupling strengths are in qualitative agreement with what we observe. Figure \ref{modeSims}a shows one free spectral range (FSR) of the measured data from Figure 2a of the main text, where we can clearly see a collection of modes that couple to the qubit. We compare this to the simulated acoustic mode spectrum shown in Figure \ref{modeSims}b, where we plot the total mode intensity versus the frequency of the acoustic excitation. The free parameters to match the simulated mode frequencies to the experimental data are the longitudinal and transverse sound velocities of sapphire, which are not well known at low temperatures. We find them to be $v_l = 11100$ m/s and $v_t = 8540$ m/s, which is similar to typical measured values and the values used in our previous work\cite{ChuScience2017SI}. All other geometrical parameters, including the shape of the curved surface, are known from independent measurements such as the one shown in Figure \ref{fab}b. The simulated spectrum agrees reasonably well with the experimental data, especially for the first few transverse modes in each FSR. There appear to be additional modes in the measured data, which may be higher order transverse modes from the neighboring longitudinal mode number. Another possibility is the existence of acoustic modes with other polarizations in the resonator, which are not included in the simulations. The qubit may couple to these modes through the non-zero transverse components of the electric field and shear components of the AlN piezoelectric tensor. 

The relative heights of the peaks in Figure \ref{modeSims}b give some sense of how well each mode is coupled to the qubit. However, to obtain the coupling rates more accurately, we find the strain profiles at only the simulated mode frequencies after $2\times10^5$ round trips. We then calculate the coupling rate, which includes the overlap of the simulated qubit electric field, the shape of the piezoelectric transducer, and the simulated acoustic mode profiles, as described in the Supplementary Materials of\cite{ChuScience2017SI}. Figure \ref{modeSims}c shows the resulting predicted coupling rate $g$ for each mode. We see that the relative coupling strengths qualitatively agree with what we experimentally observe, with the fundamental Gaussian mode being at least a factor of five more strongly coupled than any higher order mode. The overall strength of the coupling is higher in the simulations than what is observed in our experiment, which could be due to a variety of reasons. One is a lower piezoelectric constant for the actual material than the value used in the similation, which we have not independently measured. Misalignment and a larger gap between the two chips than expected would also lead to different overall and relative coupling strengths. To resolve these discrepancies, we hope to more systematically characterize the flip-chip device geometry in the future.

\section{Coherent displacement calibration}

\begin{figure}[ht] 
\begin{center}
\includegraphics[width = 0.3 \textwidth]{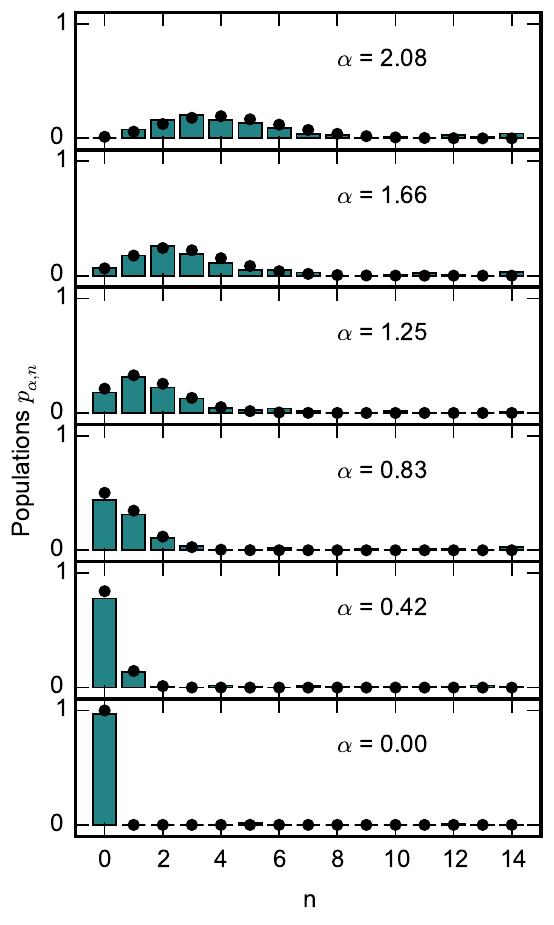}
\caption{\textbf{Coherent displacements of the phonon mode.} Bars indicate the measured populations. Black dots are the ideal Poisson distributions for each $\alpha$.
}
\label{disp}
\end{center}
\end{figure}

In this section, we show that our procedure for doing a direct displacement on the phonon mode does indeed result in a coherent state. Figure \ref{disp} shows the extracted Fock state populations after performing displacements with six different amplitudes within the range used for the measurements of Wigner functions. As in the main text, the displacement drive is a Gaussian pulse at the phonon frequency with a 1 $\mu$s RMS width and 4 $\mu$s total width. We find that they agree well with the expected Poisson distributions for coherent states whose amplitudes $\alpha$ are given by a single scale factor multiplied by the actual drive amplitudes used in the experiments. This is in practice how we calibrate the experimental drive amplitudes to the actual displacement amplitudes, and the same scale factor is used in all Wigner tomography experiments. We find that for shorter pulses and larger drive amplitudes, the population distributions deviate from that of coherent states. These effects are reproduced in simulations, and are potentially due to off-resonant excitations of the qubit. In the case of significant initial excited state population of the qubit, our procedure for measuring the phonon populations is no longer valid. On the other hand, a longer pulse with smaller amplitudes results in more decoherence during the displacement pulse, thus limiting our ability to perform Wigner tomography on a prepared state. The pulse length used here and in the data shown in Figure 4 of the main text is a compromise between these two considerations. Future optimization of the qubit detuning during the displacement and the pulse shape could potentially improve the robustness and range of amplitudes of the displacement drive.

\section{State reconstruction}
\begin{figure}[ht] 
\begin{center}
\includegraphics[width = 0.7 \textwidth]{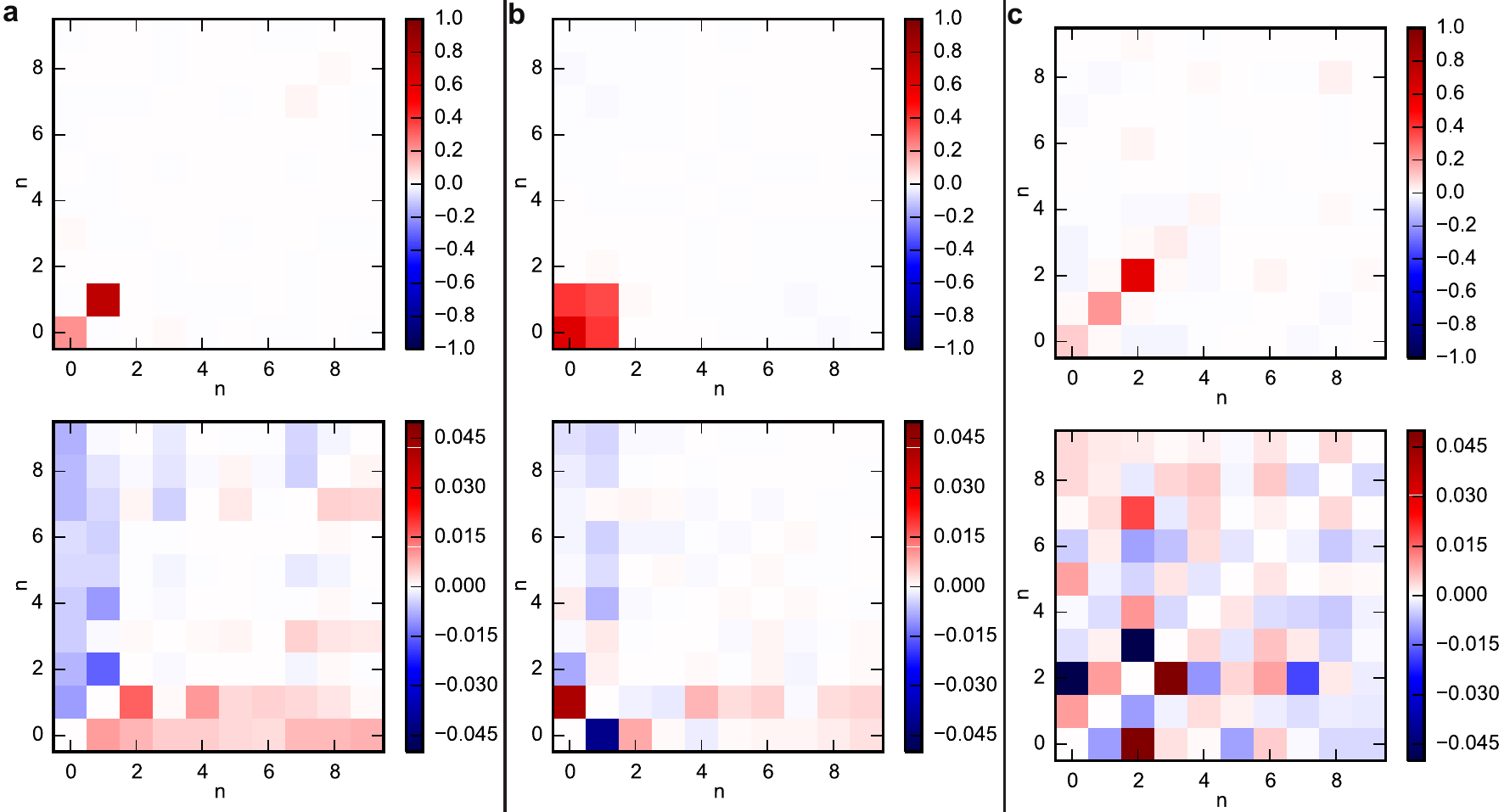}
\caption{\textbf{Density matricies.} Real (top row) and imaginary (bottom row) parts of the density matrices for the states a. $\ket{1}$, b. $(\ket{0}+\ket{1})/\sqrt{2}$, and c. $\ket{2}$.
}
\label{rho}
\end{center}
\end{figure}

State reconstruction was done using the formalism described in\cite{ChouArxiv2018SI}. We use a maximum likelihood method that finds the most probable density matrix $\rho$ that results in the Wigner functions we measured, subject to the constraints that $\rho$ is physical (positive semi-definite, Hermitian, and unit trace). We used Fock states up to $n=14$ for extracting the populations $p_{n, \rho}(\alpha)$ of both the Fock states shown in Figure 3 of the main text and for calculating the Wigner functions shown in Figure 4 of the main text. We truncate the Hilbert space at Fock state $n=9$ for the reconstruction. Figure \ref{rho} shows the reconstructed density matrices that were used to plot the Wigner functions in Figures 4d, e, and f of the main text.

\end{document}